\documentclass[prl,a4paper,superscriptaddress,floatfix,notitlepage]{revtex4}
\PassOptionsToPackage{usenames,dvipsnames}{color}
\usepackage{amsmath,amsfonts,amssymb,amsthm,graphicx,bbm,enumerate,times,color,mathrsfs}
\usepackage{dsfont}
\usepackage{times}
\usepackage[normalem]{ulem}
\usepackage{mathtools}
\usepackage[makeroom]{cancel}

\usepackage{natbib}
\setcitestyle{square,comma,numbers}
 \usepackage{amsfonts}
\usepackage{amsmath}
 \usepackage{color}
 \usepackage{amssymb}

\usepackage[colorlinks=true,urlcolor=blue]{hyperref}

\newcommand{\id}{\mathbb{I}}
\DeclarePairedDelimiter\ceil{\lceil}{\rceil}
\DeclarePairedDelimiter\floor{\lfloor}{\rfloor}

\newcommand{\ket}[1]{|#1\rangle}
\newcommand{\bra}[1]{\langle #1|}

\begin{document}
\title{Collective operations can extremely reduce work fluctuations }

\author{Mart\'i~Perarnau-Llobet}
\address{Max-Planck-Institut f\"ur Quantenoptik, Hans-Kopfermann-Str. 1, D-85748 Garching, Germany}

\author{Raam Uzdin}
\address{Schulich Faculty of Chemistry, Technion – Israel Institute of Technology, Haifa 3200000, Israel}

\begin{abstract}
We consider work extraction from $N$ copies of a quantum system. 
When the same work-extraction process is implemented on each copy, the relative size of fluctuations is expected to decay as $1/\sqrt{N}$. 
Here, we consider protocols where  the copies can be processed collectively, and  show that in this case work fluctuations can disappear exponentially fast in $N$. As a consequence, a considerable proportion of the average extractable work $\mathcal{W}$ can be obtained  almost deterministically by globally processing a few copies of the state. 
This  is derived in the two canonical scenarios for work extraction: (i)~in thermally isolated systems, where $\mathcal{W}$ corresponds to the energy difference between  initial and  passive  states, known as the ergotropy, and (ii)~in the presence of a thermal bath, where $\mathcal{W}$ is given by the free energy difference between  initial  and  thermal states.
\end{abstract}

\maketitle

\section{Introduction}
Fluctuations play a crucial role in microscopic systems, which has
motivated enormous efforts to build a theory of thermodynamics for
small fluctuating systems, both classical~\cite{Jarzynskia2008}
and quantum~\cite{Esposito2009nonequilibrium,Campisi2011Colloquim,Hanggi2015theother}.
Particularly relevant is the question of how to extend the celebrated
second law of thermodynamics. A standard formulation of this law is
that the average work consumed or extracted from a system in a process
is bounded by its change of free energy. Since this law concerns average
work, one can naturally ask what can be said about its fluctuations.
This question has lead to very interesting insights, which notably
include fluctuation theorems~\cite{Hanggi2015theother} and extensions
of the free energy to the nanoscale regime~\cite{horodecki2013fundamental,aaberg2013trulyNC}.
This rich behaviour is smoothed out in the macroscopic limit, where
fluctuations can usually be neglected with respect to average quantities.
To formalise this limit, consider $N$ identical copies of a quantum
system, $\rho^{\otimes N}$, and assume that the same work extraction
process is implemented on each~$\rho$. Then, the law of large numbers
ensures that work fluctuations have a size proportional to $\sqrt{N}$,
whereas average work scales as $N$. The relative size of fluctuations
will hence decay as $1/\sqrt{N}$, regardless of the specific protocol
being implemented. The universality of this behaviour builds a clear
intuition that (work) fluctuations disappear in the limit $1/\sqrt{N}\rightarrow0$.

In this article, we also consider work extraction from $\rho^{\otimes N}$,
but away from the thermodynamic limit $1/\sqrt{N}\rightarrow0$, so
that $N$ can be small. Crucially, we consider processes where the
$N$ copies of $\rho$ can be processed collectively, hence assuming
a high level of control. In this case, using concentration results
from probability and information theory~\cite{Shannon1948,Hoeffding1963},
we find that work fluctuations can disappear exponentially fast in~$N$.
This is first derived in thermally isolated systems, where the amount
of extractable work from a state $\rho$ is given by its ergotropy
$\mathcal{W}_{{\rm erg}}(\rho)$~\cite{Allahverdyan2004}, i.e.,
the energy difference between $\rho$ and its passive state~\cite{Pusz1978,Lenard1978}.
While the concepts of ergotropy and passivity have been explored for
average quantities~\cite{Pusz1978,Lenard1978,Allahverdyan2004,Skrzypczyk2015,meps2015,sparaciari2017energetic},
our results show that a large proportion of $\mathcal{W}_{{\rm erg}}(\rho^{\otimes N})$
can be extracted \emph{without fluctuations} by means of global operations.
We note that the relevance of work fluctuations has been characterised
in the creation of correlations from work~\cite{Nayeli2018}, and
in the charging of a quantum battery consisting of an harmonic oscillator~\cite{Friis2018precisionandwork}.
We also study the decay of work fluctuations in the presence of a
thermal bath. In this case, fluctuation-free protocols have been extensively
studied in the last years within the field of single-shot thermodynamics~\cite{Dahlsten2011,brandao:2013,horodecki2013fundamental,aaberg2013trulyNC,Halpern2015,Gemmer2015,richens2016work,renes2016relative,VanDerMeer2017,chubb2017beyond,Faist2018},
and our results provide new insights on the mesoscopic regime~\cite{VanDerMeer2017,chubb2017beyond,Tajima2017,Ito2018,Richens2018,Scharlau2018},
where $N$ is finite and possibly small. Our considerations concern
states that are diagonal in the energy basis, as the definition of
work fluctuations in coherent systems is subtle and an active area
of research~\cite{baumer2018fluctuating}.

\section{Work extraction from N two-level systems}

 We start by considering
an illustrative scenario where work is extracted from a system S made
up of $N$ identical qubits (spins). 
 This is modelled by putting S in contact with a work repository
W, which can extract or give energy to S. The total Hamiltonian of
SW reads $H=H_{S}+H_{W}$, where $H_{S}=\sum_{i=1}^{N}\id^{\otimes i-1}\otimes h\otimes\id^{\otimes N-i+1}$
with $h=\nu\ket 1\bra 1$, and $H_{W}=\sum_{j=-L}^{L}w_{j}\ket{w_{j}}\bra{w_{j}}$
with $w_{j}=j\nu$ and $L\geq N$. Energy is transferred from S to
W via unitary operations that satisfy 
\begin{equation}
[U,H]=0,\label{energyconsII}
\end{equation}
which ensures that energy extracted/consumed from $S$ comes solely
from $W$. The initial state of $SW$ is taken to be a product state,
\begin{equation}
\sigma=\rho^{\otimes N}\otimes\ket 0\bra 0,\label{sigma1}
\end{equation}
with $\rho=(1-p)\ket 0\bra 0+p\ket 1\bra 1$ being the single-qubit
density matrix. We also assume $p>1/2$ so that the qubits have initially
population inversions and can transfer energy to W.

Work is quantified as energy exchanges in the state of W, which starts
with a well-defined energy \cite{Note5}. Given (\ref{sigma1}), the
probability to extract an amount of work $w$ through $U$ reads $P(w)=\bra w{\rm Tr}_{S}\left(U\sigma U^{\dagger}\right)\ket w$,
i.e., the probability that $W$ has raised from $0$ to $w$. Our
goal is to maximise $P(w)$ over all protocols, and hence we define
\begin{equation}
\tilde{P}(w)=\max_{U}\left(\bra w{\rm Tr}_{S}\left(U\sigma U^{\dagger}\right)\ket w\right),\label{PsuccII}
\end{equation}
where the maximisation runs over $U$'s that satisfy (\ref{energyconsII}). Note that in principle the optimal $U$ depends on $w$. 
To compute of (\ref{PsuccII}), we take 
$w=k\nu$ and consider transitions in degenerate energy levels of
SW, which is imposed by~(\ref{energyconsII}). Considering a global
energy $E_{SW}=j\nu$ (with $j\in \{0,...,N\}$) the populations of $SW$
are distributed as: 
\begin{itemize}
\item Initial state: $W$ is at $\ket 0$. Then there are $C_{N}^{j}\equiv{N \choose j}$
states of SW at energy $E_{SW}=j\nu$, each with probability $p(j)\equiv p^{j}(1-p)^{j}$.
\item Target state: $W$ is at $\ket{k\nu}$, and there are $C_{N}^{j-k}$
states at energy $E_{SW}=j\nu$. Note that initially these states
are not populated. 
\end{itemize}
The protocol that achieves $\tilde{P}(k\nu)$ is the one that moves
as much probability as possible from the initial to the target state
in each degenerate energy $E_{{\rm SW}}=j\nu$. Since we only consider
diagonal states, it is enough to consider permutations within each
subspace that move the highest populations to the desired state (note that creating coherence can only mix the probabilities, and hence it cannot increase the population of the target state). 
This leads to
\begin{eqnarray}
\tilde{P}(k\nu)& = \sum_{j=k}^{N}\min\left(C_{N}^{j-k},C_{N}^{j}\right)p(j)
\nonumber\\
  & =  \sum_{j=k}^{\ceil{(N+k)/2}-1}C_{N}^{j-k}p^{j}(1-p)^{N-j}+
    \sum_{j=\ceil{(N+k)/2}}^{N}C_{N}^{j}p^{j}(1-p)^{N-j}
    \label{eq: P_exact_binom}
\end{eqnarray}
The first term is positive and small when $\ceil{(N+k)/2}\ll Np$.
By removing it we obtain
\begin{eqnarray}
\tilde{P}(k\nu) & \geq  \sum_{j=\ceil{(N+k)/2}}^{N}C_{N}^{j}p^{j}(1-p)^{N-j}
\nonumber\\
&=1-\sum_{j=0}^{\ceil{(N+k)/2}-1}\hspace{-1mm}C_{N}^{j}p^{j}(1-p)^{N-j}.
\end{eqnarray}
Defining the single-qubit ergotropy~\cite{Allahverdyan2004} (see also \eqref{ergotropy} for a definition of the ergotropy)
\begin{equation}
\mathcal{W}_{{\rm qbit}}\equiv(2p-1)\nu,
\label{Wqubits}
\end{equation}
we  focus on the case $k\nu=N\mathcal{W}_{{\rm qbit}}(1-\gamma)$,
with $\mathcal{W}_{{\rm qbit}}\equiv(2p-1)\nu$ so that $k=N(2p-1)(1-\gamma)$
with $\gamma\in(0,1)$. We then use Hoeffding's inequality \cite{Hoeffding1963},
which in the particular case of a Binomial distribution with probability
of success $p$ and $N$ runs, states that the probability $p(l)$
of obtaining $l\leq Np$ successes is bounded by $p(k)\leq\exp(-2(Np-l)^{2}/N)$.
Here, this implies 
\begin{eqnarray}
\tilde{P}(k\nu) & \geq & 1-\exp\left(-2N\left(p-\frac{\ceil{(N+k)/2}+1}{N}\right)^{2}\right)\\
 & \geq & 1-\exp\left(-N\gamma^{2}2(p-1/2)^{2}\right)\label{eq:quadraticbound}
\end{eqnarray}
where the second inequality follows from$\ceil{(N+k)/2}\le(N+k)/2+1$.
An tighter bound can be obtained through the relative entropy bound on the binomial tail $P(l)\le\exp\left(-ND(\frac{l}{N}||p\right)$
where $D(x||y)=x\ln\frac{x}{y}+(1-x)\ln\frac{1-x}{1-y}$, which leads to
\begin{eqnarray}
\tilde{P}(k\nu) & \geq & 1-\exp\left(-ND(\frac{\ceil{(N+k)/2}+1}{N}||p)\right)\\
 & \geq & 1-\exp\left(-ND(\frac{1}{2}+\frac{k}{2N}||p)\right),\label{eq:rel ent bound}
\end{eqnarray}
where once again the second inequality follows from $\ceil{(N+k)/2}\le(N+k)/2+1$
when $\frac{1}{2}+\frac{\ceil{(N+k)/2}+1}{N}\le p$ (which is equivalent
to the sub-ergotropy work extraction $\gamma<0$). 
To interpret these result, note that $\mathcal{W}_{{\rm qbit}}$ in
(\ref{Wqubits}) corresponds to the extractable work \emph{on average}
from $\rho$, i.e., its ergotropy \cite{Allahverdyan2004} ( see also
discussion in (\ref{ergotropy})). Hence, this result shows that we
can extract deterministically a proportion $1-\gamma$ of the total
\emph{average} extractable work from $\rho^{\otimes N}$ with a failure
probability that decays exponentially with $\gamma^{2}N$.

\begin{figure}[htp]
\centering
\includegraphics[scale=0.4]{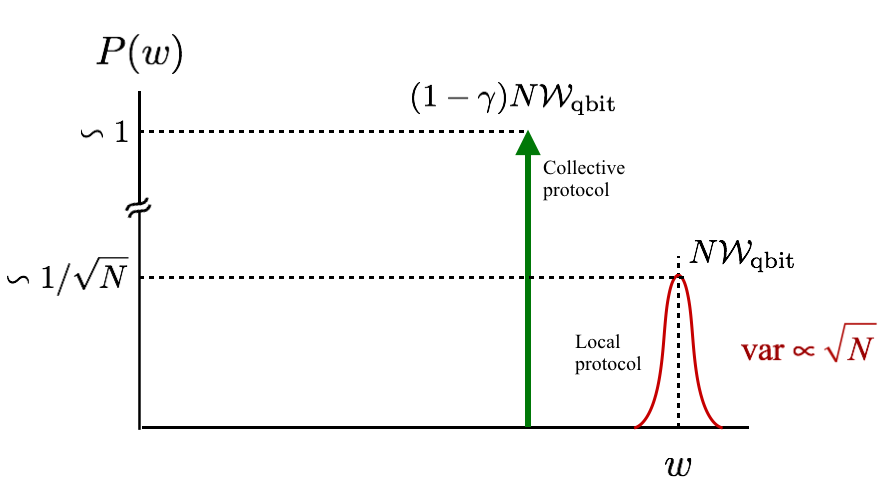}
\caption{\label{fig:R} Schematic illustration of the difference between local
and global protocols for work extraction of $N$ qubits with excitation
probability $p$ and $\nu=1$. Whereas local protocols always have
fluctuations $\propto\sqrt{N}$ which diminish the success probability
$P(w)$, global protocols can exponentially supress by sacrificing a
small portion $\gamma$ of the average available work (ergotropy).}
\end{figure}



These results are a direct consequence of concentration results in probability theory, in particular of Hoeffding's inequality. Yet, it is important to distinguish it from the standard thermodynamic limit in which the relative size of fluctuations disappears as $1/\sqrt{N}$. To illustrate this point, consider a simple protocol where work is extracted from each qubit individually. Energy is transferred from S to W by iteratively applying   the energy-preserving unitary that swaps $\ket{0,j+1} \leftrightarrow \ket{1,j}$ $\forall j$,
 where $\ket{a,b}\equiv \ket{a}_S\otimes \ket{b}_W$, successively for  each of the $N$ copies. After $N$ iterations, the state of $W$ is described by  $\sigma_W= \sum_{j=0}^{N}C^j_N p(j) \ket{2j-N} \bra{2j-N}.$
The work distribution $P(w)=\bra{w} \sigma_W \ket{w}$ is hence a binomial distribution centred at $\langle w \rangle =N(2p-1)\nu$ with standard deviation $\mu=\nu \sqrt{Np(1-p)}$. 
 Hence, this protocol is essentially probabilistic and becomes only reliable by considering a confindence interval $N\mathcal{W}_{\rm qbit}\pm \mathcal{O}(\sqrt{N})$ \cite{Note1}. This is  contrast with the  fast convergence to a deterministic process  of global protocols suggested by \eqref{eq:quadraticbound}. This difference appears in protocols that extract less than $N\mathcal{W}_{\rm qbit}$.

\begin{figure}[htp]
\centering
\includegraphics[scale=0.3]{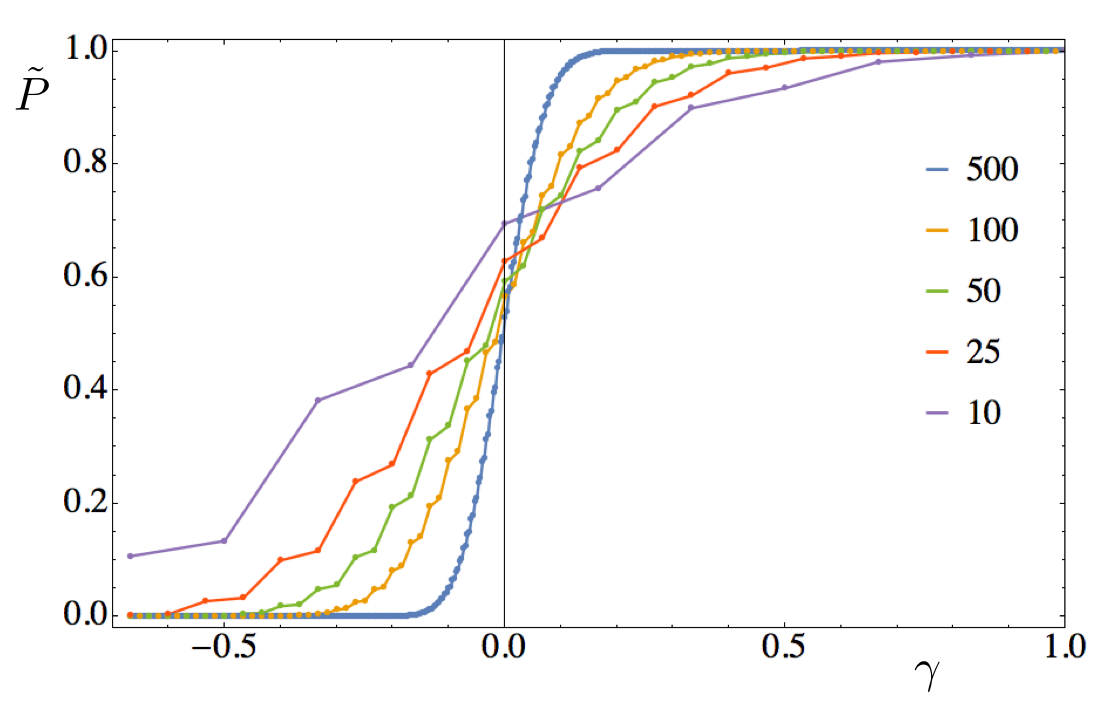}
\caption{\label{fig:resultx}   
Maximal success probability $\tilde{P}(\omega_{\gamma})$ as a function of the deviation from the ergotropy $\gamma$, with $\omega_{\gamma}=(1+\gamma)N\mathcal{W}_{\rm erg}(\rho)$ and where $\mathcal{W}_{\rm erg}(\rho)$ is the ergotropy given in \eqref{Wqubits}. Parameters: $p=0.8$.   
}
\end{figure}

This simple example demonstrates the capabilities of global protocols for reducing work fluctuations.
This is  illustrated in Fig. \ref{fig:resultx}, where we plot $\tilde{P}(\omega_{\gamma})$, with $\omega_{\gamma}= N\mathcal{W}_{\rm qbit} (1-\gamma)$, as a function of $\gamma$ and for different  $N$'s. Note, for example, that one can extract 2/3 of the average extractable work from $\rho^{\otimes N}$ with success probability  $\tilde{P}\approx \{0.90,0.92,0.97,0.99 \}$   by collectively processing $N=\{10,25,50,100 \}$ copies of the state. On the other hand, the success probability of protocols that attempt to extract more than $N \mathcal{W}_{\rm qbit}$ rapidly decays to zero. 

\begin{figure}
\centering
\includegraphics[scale=0.75]{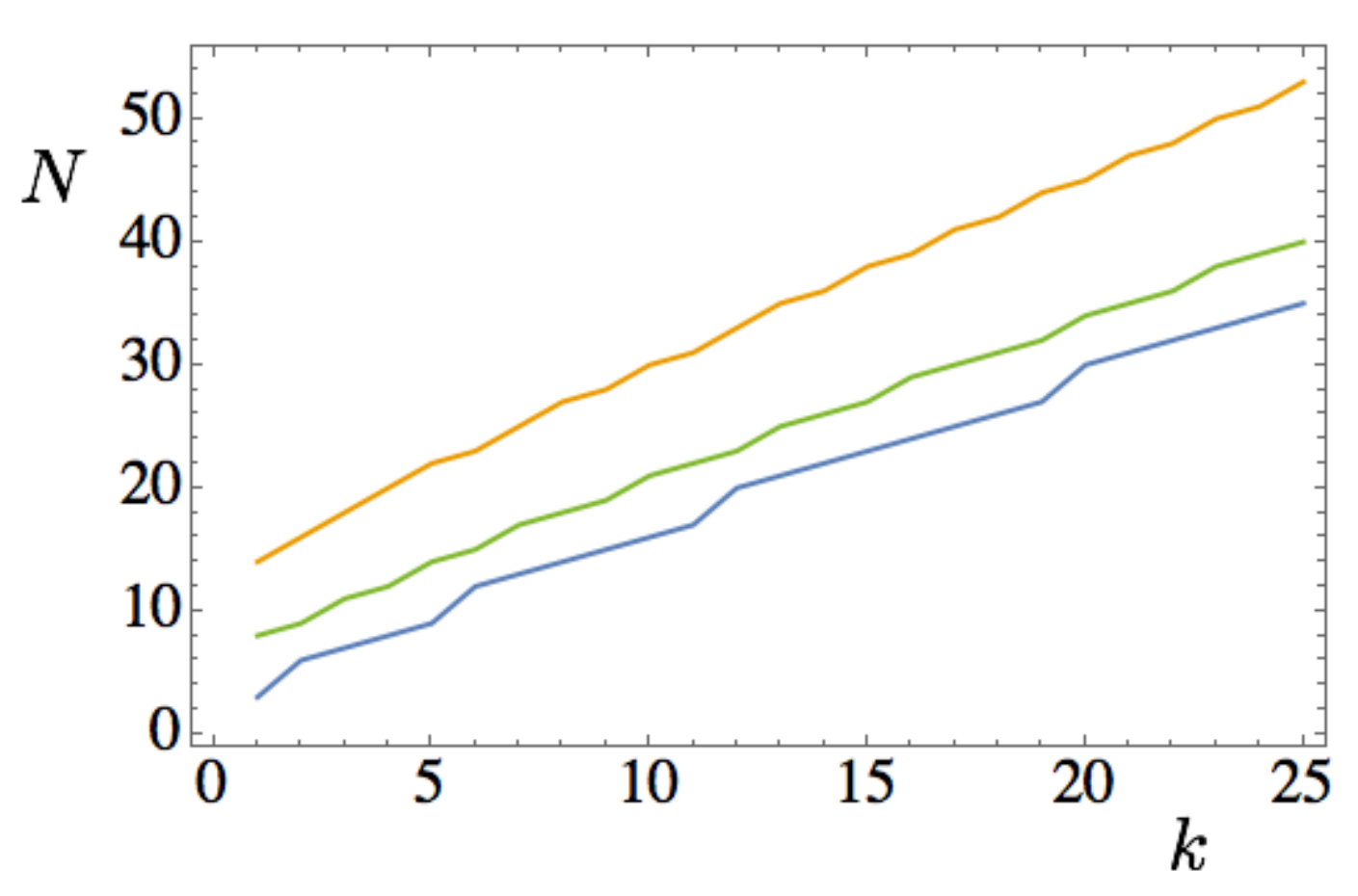}
\caption{\label{fig: N comp}The exact minimal number of spins with population
inversion ($p=0.95$) needed to extract different values of work  (computed through $w=k\nu$) with
success probability of 0.99 is shown in blue. The green (blue) curve
show our relative entropy (quadratic) bound.}
\end{figure}

Applying global operations can be quite challenging as it involves
synchronized interaction of many particles. Imagine a scenario where
the number of population inverted spins $N_{tot}$ is of the order
of thousands. The goal is to extract $k_{tot}\nu$ work with some
success probability $P_{0}$ ($P_{0}$ is very close to one). Doing
a global operation on all of them together would be optimal in terms
of the minimisation of fluctuations, but it would more practical to divide the total number
spin to sets of about dozens each, and apply a simpler global operation
multiple on each set. Thus, it is useful to evaluate what is the smallest
number of spins needed for achieving the success probability goal.
Settting  $D(\frac{1}{2}+\frac{k}{2N},p)\to D(\frac{1}{2}+\frac{k_{tot}}{2N_{tot}},p)$ in (\ref{eq:rel ent bound}), we obtain that for unit size of 
\begin{equation}
N\ge \frac{\ln\left(\frac{1}{1-P_{0}}\right)}{D\left(\frac{1}{2}+\frac{k_{tot}}{2N_{tot}},p\right)},\label{eq: N est}
\end{equation}
 $\frac{k_{tot}}{N_{tot}}N\nu$ work extraction is guaranteed with
success probability of at least $P_{0}$. As an example we consider
spins with probability $p=0.95$ to be in the excited state. The blue
curve in Fig. \ref{fig: N comp} shows the exact number of spins needed
to guarantee a success probability of $0.99$, as a function of the
extracted work. The red curve shows the estimation of $N$ using (\ref{eq:quadraticbound})
and the relative entropy bound (\ref{eq: N est}) is shown in green.

\section{Disappearance of work fluctuations in thermally isolated systems.}

In this section we show that the behaviour explained in the previous section for a collection of spins is in fact  a generic property of work-extraction protocols. In what follows, we prove that 
\begin{align}
\tilde{P}\big(\omega_{-\gamma}\big)\geq 1-d\exp\left(-N \gamma^2 c^2 \right),
\label{boundII}
\end{align}
with 
\begin{align}
\omega_{-\gamma}=(1-\gamma)N \mathcal{W},
\label{omegagamma}
\end{align}
 where  $\gamma \in (0,1)$ and  $\mathcal{W}$ is the maximal extractable work \emph{on average} from a  diagonal state $\rho$ (i.e., its ergotropy); whereas $c$ and $d$ in \eqref{boundII}  are parameters that depend on the initial conditions  (the initial state $\rho$ and Hamiltonian $H_S$).  Expressions \eqref{boundII} and \eqref{omegagamma} naturally connect average bounds with almost-deterministic protocols for work extraction.  
  A direct consequence of this result is to provide a bound on the minimal number of copies that we need to globally process in order to extract $\omega_{-\gamma}$ with success probability $1-\epsilon$, $N \geq \ln(d/\epsilon)/(\gamma^2 c^2).$
  
 Consider  that S is  made up of $N$ identical ($d$-level) qudit systems, each of them with a Hamiltonian $h= \sum_{i=1}^d \nu_i \ket{i}\bra{i}$ and an initial diagonal state  $\rho=\sum_{i=1}^d p_i \ket{i}\bra{i}$. The total state reads $\rho^{\otimes N}$ and the total Hamiltonian is non-interacting, $H_S= \sum_{i=1}^{N} \id^{\otimes i-1} \otimes h \otimes \id^{\otimes N-i+1}$. The state is thermally isolated, so that work can only be extracted via controlled operations. Then, a crucial quantity is the ergotropy of $\rho$~\cite{Allahverdyan2004}
 \begin{align}
\mathcal{W}_{\rm erg}(\rho) \equiv  {\rm Tr}(h(\rho - \rho_{\rm pas})),
 \label{ergotropy}
 \end{align}
 where   $\rho_{\rm pas}$ is the passive state  associated to $\rho$ \cite{Pusz1978,Lenard1978}, 
  \begin{align}
 \rho_{\rm pas}=\sum_i p_i^{\downarrow} \ket{i}\bra{i},
 \end{align}
 with $p_i^{\downarrow}$ being the eigenvalues of $\rho$ ordered in decreasing order. $\mathcal{W}_{\rm erg}(\rho)$ quantifies the average extractable work from $\rho$ by means of unitary operations as ${\rm Tr}(h(\rho-U\rho U^{\dagger})\leq \mathcal{W}_{\rm erg}(\rho)$, $\forall U$ \cite{Note2}.  That is, it quantifies how much accessible energy  on average is stored in $\rho$.
 
 While the ergotropy  defined in ~\cite{Allahverdyan2004} deals with average work, here we aim to consider probabilistic work-extraction protocols.  For that,  as in the previous section, we consider an  auxiliary work-repository W, which can accept energy from S.  The possible work values are given by energy differences of $H_S$,  which can be written as
 \begin{align}
 w_{{\bf n},{\bf m}} = \sum_{i=1}^d (n_i-m_i) \nu_i
 \label{valueswork}
 \end{align}
 with ${\bf n}\equiv (n_1,n_2,...,n_d)$,   $n_i \in [0,N]$, $\sum_i n_i=N$ and similarly for ${\bf m}$. The spectrum of W is taken to be non-degenerate and dense enough to accept all possible $w_{{\bf n},{\bf m}}$. 
 
 \begin{figure}[htp]
 \centering
\includegraphics[scale=0.3]{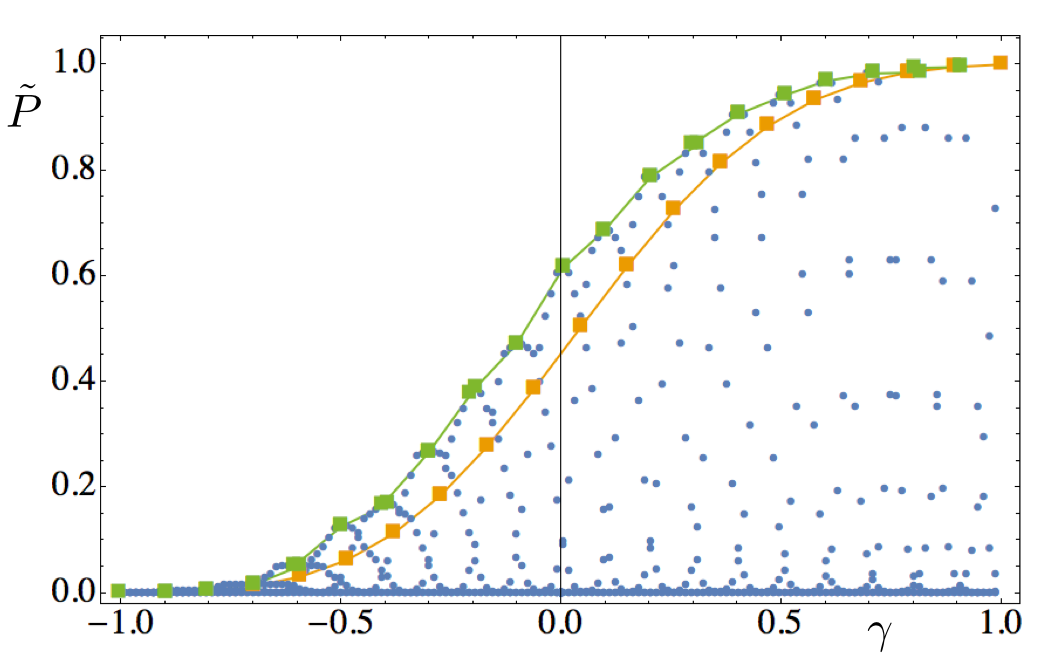}
\caption{\label{fig2} $\tilde{P}(\omega_{\gamma})$  as a function of $\gamma$, with $\omega_{\gamma}=(1+\gamma)\mathcal{W}_{\rm erg}^{\rm glob}(\rho)$ for 20 copies of a qutrit system $\rho$ with populations $\{p_0,p_1,p_2\} =\{0.2,0.2,0.6\}$ and internal Hamiltonian $h=\{0,1,8 \}$. The blue dots correspond to  $ \tilde{P}(w_{{\bf n},{\bf m}})$ $\forall {\bf n}, {\bf m}$, whereas the orange and green dots correspond to the choices \eqref{n_ikl} and \eqref{n_iklII}, respectively. 
 }
\end{figure}
 
 We now proceed to find a lower bound on $\tilde{P}(w)$ as defined in  \eqref{PsuccII}, where the maximisation runs over all unitaries satisfying \eqref{energyconsII}. The possible values of $w$ are given by \eqref{valueswork} with all possible $w_{{\bf n},{\bf m}}$. We focus on a subset of those, defined by   
 \begin{align}
 n_i-m_i =\lfloor N (1-\gamma) (p_i-p_i^{\downarrow})\rfloor\equiv k_i,
 \label{n_ikl}
 \end{align}
 where $\gamma \in (0,1) $, and the $\lfloor...\rfloor$ ensures that the $k_i$ are natural numbers. This can be understood as a  work-extraction protocol which moves the probability of the S' initial states  with energy $\sum_i j_i \nu_i $ to target states of S with energy $\sum_i (j_i-k_i) \nu_i $,~$\forall{{\bf j}}$. As a consequence, W raises from $0$ to   $w=\sum_i k_i\nu_i \approx (1-\gamma)N \mathcal{W}_{\rm erg}(\rho)$. Given this protocol, the calculation of $\tilde{P}(w)$ is conceptually similar to the previous qubit example,  but becomes considerably more involved and is carried out in \ref{section.qdit}. 
 There, using standard concentration techniques in information theory~\cite{Shannon1948,Hoeffding1963,wilde2013quantum}),  we show that \eqref{boundII}  and \eqref{omegagamma} are satisfied with $\mathcal{W}=\mathcal{W}_{\rm erg}(\rho)$, $d=d_S$ where $d_S$ is the dimension of $h$, and where $c$, at leading order in $1/N$ and $\gamma$,  is given by 
 \begin{align}
c=\frac{\sqrt{2}\hspace{0.5mm}S(\rho_{\rm pas}||\rho)}{\sum_{i=1}^{\lfloor d/2 \rfloor} \ln h_i^{\downarrow} - \sum_{i=\lceil d/2 \rceil+1}^{d} \ln h_i^{\downarrow} }
\label{gqudits}
 \end{align}
where $h_i\equiv p_i^{\downarrow}/p_i$ and where $S(\rho||\sigma)$ is the relative entropy. 
We stress that  the expression \eqref{gqudits} for $c$ is correct  up to corrections of order $\mathcal{O}(1/N)$ and $\mathcal{O}(\gamma^2)$, the first order corrections in $\mathcal{O}(1/N)$ is provided in \ref{section.qdit}. 

An exciting property of passive states is the phenomenon of \emph{activation}~\cite{Pusz1978,Lenard1978,Alicki2013,Skrzypczyk2015,meps2015,sparaciari2017energetic}, which means that $\mathcal{W}_{\rm erg}(\rho^{\otimes N})$ can be larger than $N\mathcal{W}_{\rm erg}(\rho)$. That is, by collectively processing $\rho^{\otimes N}$ more  \emph{average} work can be extracted. Interestingly, Alicki and Fannes showed in \cite{Alicki2013}  that for $N\rightarrow \infty$, 
\begin{align}
 \mathcal{W}_{\rm erg}^{\rm glob}(\rho) \equiv \hspace{-1.5mm}\lim_{N \rightarrow \infty} \frac{\mathcal{W}_{\rm erg}(\rho^{\otimes N})}{N}= {\rm Tr}(h(\rho-\omega_{\beta}(h))),
\label{thermoBound}
\end{align}
where $\omega_{\beta}(h)$ is a thermal state, $\omega_{\beta}(h)\equiv e^{-\beta H}/{\rm Tr}(e^{-\beta H})$, whose temperature is defined by $S(\rho)=S(\omega_{\beta}(h))$ with $S(\rho)=-{\rm Tr}(\rho \ln \rho)$ the Von Neumann entropy \cite{Note3}. 
Note   that \eqref{thermoBound} can be also obtained through catalytic transformations \cite{sparaciari2017energetic}, and that the state after the transformation is a completely passive state, i.e., a Gibbs state, which is useless for work extraction \cite{Pusz1978,Lenard1978,Skrzypczyk2015}. 

A natural question then arises: Can one extract a large proportion of the fundamental bound \eqref{thermoBound} almost deterministically? In \ref{section.fluctuations.activation}, we answer this question positively. 
More precisely, we consider a  protocol defined by 
 \begin{align}
 n_i-m_i=\lfloor N (1-\gamma) (p_i-p_i^{\rm th}) \rfloor \equiv k_i^{\rm th},
 \label{n_iklII}
 \end{align}
 where $\gamma \in (0,1) $,  and $p_i^{\rm th}$ are the populations of $\omega_{\beta}(h)$, i.e., $\omega_{\beta}(h)=\sum_i p_i^{\rm th}\ket{i}\bra{i}$.
For such $k_i^{\rm th}$'s, we then show that  \eqref{boundII}  and \eqref{omegagamma} are satisfied with $\mathcal{W}=\mathcal{W}_{\rm erg}^{\rm glob}(\rho)$, $d=d_S$, and where $c$ can be inferred from \eqref{gqudits} by replacing $\rho_{\rm pas} \leftrightarrow \omega_{\beta}(h)$ and $h_i \leftrightarrow h^{\rm th.}_i \equiv p_i^{\rm th}/p_i$.  This value of $c$ holds up to corrections $\mathcal{O}(1/N)$, which are expected to appear as \eqref{thermoBound} can only be achieved for large $N$.

We illustrate these results in Fig. \ref{fig2}, where we compute exactly $\tilde{P}(w)$ for 50 copies of a qutrit system. From the figure it becomes clear that there are many possible protocols corresponding to different choices of $ w_{{\bf n},{\bf m}}$, and that the  choices \eqref{n_ikl} and \eqref{n_iklII} perform particularly well, especially the latter. In fact, numerical results suggest that  \eqref{n_iklII} rapidly becomes optimal with increasing $N$.

\section{Disappearance of work fluctuations for systems in contact with a thermal bath} We now move to the case where S may be put in contact with a thermal bath B at temperature $1/\beta$.  
In this case, the average extractable work from $\rho$ with internal Hamiltonian $h$  is given by the  non-equilibrium free energy change~\cite{procaccia1976potential,takara2010generalization,Alicki2004,Esposito2011,Popescu2013},
\begin{align}
\mathcal{W}_{\rm th}(\rho)= F(\rho,h)-F(\omega_{\beta}(h),h),
\label{freeenergydif}
\end{align} 
where $F(\rho,h)={\rm Tr}(\rho h)-T S(\rho)$, with $S(\rho)=-{\rm Tr}(\rho \ln \rho)$ and $\omega_{\beta}(h)=e^{-\beta h}/{\rm Tr}(e^{-\beta h})$,  the temperature being now provided by B. 
The  expression \eqref{freeenergydif} bounds the extractable work on average of a generic $\rho$ with B, and can be interpreted as a formulation of   the second law of thermodynamics with a single bath. Note also that $N\mathcal{W}_{\rm th}(\rho)=\mathcal{W}_{\rm th}(\rho^{\otimes N})$ when the total Hamiltonian $H_S$ is a sum of non-interacting $h$'s, so there is no possible activation here. 

Now we move to probabilistic work-extraction protocols on $\rho^{\otimes N}$ in contact with B. We consider $\rho=\sum_{i=1}^d p_i \ket{i}\bra{i}$, i.e. diagonal states,  with internal Hamiltonian $h= \sum_{i=1}^d \nu_i \ket{i}\bra{i}$ with $\nu_1=0$  for convenience.  The bath has an internal Hamiltonian $H_B$ and starts in a Gibbs state, $\omega_{\beta}(H_B)$ --see \ref{section.qdit.bath} for details on the spectra  of B. The total Hamiltonian is then $H=H_S+H_B+H_W$, and the initial state reads  $\sigma=\rho^{\otimes N}  \otimes \omega_{\beta}(H_B) \otimes \ket{0}\bra{0}$. Work is extracted by raising W from $0$ to $w$ through unitaries that preserve the total energy, $[H,U]=0$. This  framework  is commonly employed in resource-theoretic  \cite{janzing2000thermodynamic,brandao:2013,gour2015resource} and single-shot ($N=1$)  \cite{Dahlsten2011,brandao:2013,horodecki2013fundamental,aaberg2013trulyNC,Halpern2015,
Gemmer2015,richens2016work,renes2016relative,Lostaglio2015,VanDerMeer2017,
chubb2017beyond,Faist2018} studies of thermodynamics.

Our goal is to find $\tilde{P}(w)$ in \eqref{PsuccII}, and in particular to prove~\eqref{boundII}.  
In \ref{section.qdit.bath},  we bound $\tilde{P}(w)$ in \eqref{PsuccII} by combining the techniques developed in Ref.\cite{horodecki2013fundamental} with notions of typicality and concentration results in probability theory \cite{wilde2013quantum}. This leads to the desired results  \eqref{boundII}  and \eqref{omegagamma} with $\mathcal{W}=\mathcal{W}_{\rm th}(\rho)$ and
\begin{align}
c=\frac{S(\rho||\omega_{\beta}(h))}{\beta \sum_i \nu_i -\sum_{i} \ln p_i}.
\label{cbath}
\end{align}
where $S(\rho||\omega_{\beta}(h))$ is the relative entropy,  that satisfies $S(\rho||\omega_{\beta}(h))=\beta \mathcal{W}_{\rm th}(\rho)$. Note that  $c>0$ and that this result  holds for every finite $N$.

Our results hence show that by sacrificing a proportion $(1-\gamma)$ of the extracted work allows for \emph{exponentially} reducing its fluctuations. This is  possible through collective operations.  To give a feeling of these results, in \ref{appnumerics} we compute exactly $\tilde{P}(w)$ for a qubit system, which in particular shows that one can extract $2/3$ of the free energy of $\rho^{\otimes N}$ with a success probability $\{0.92,0.96,0.98,1.0\}$ by processing $\rho^{\otimes N}$ collectively with $N=\{10,25,50,100 \}$ (see \ref{appnumerics} for more results).

While so far we have assumed $\gamma$ to be a finite fixed number, let us now consider the limit $\gamma\overset{N\rightarrow\infty}{\longrightarrow} 0$.  In particular,  by choosing $\gamma^2=\ln(N)/(c^2N)$, we obtain simultaneously: 
\begin{align}
&\tilde{P}(w)\geq 1-d/N,
\nonumber\\
&w=\left(1-\frac{1}{c}\sqrt{\frac{\ln N}{N}} \right)N\mathcal{W}_{\rm th}(\rho),
\label{convergence}
\end{align}
Hence both $\tilde{P}(w) \rightarrow 0$ and $w \rightarrow N \mathcal{W}_{\rm th}(\rho)$ in the thermodynamic limit $1/\sqrt{N} \rightarrow 0$. Alternatively, one may choose  $\gamma^2=-\ln({\epsilon}/d)/(c^2N)$, so that one allows a finite error $\epsilon$ while $w=N\mathcal{W}_{\rm th}(\rho)+\mathcal{O}(1/\sqrt{N})$. With this choice, we recover the same behaviour  found  within single-shot thermodynamics  \cite{brandao:2013,horodecki2013fundamental,aaberg2013trulyNC,gour2015resource,chubb2017beyond}, where the usual   mindset is to allow a small finite error $\epsilon$ in order to extract $w$ exactly.  
Summarising, by appropriately choosing $\gamma$ as a function of $N$, one can interpolate between two regimes of work extraction from $\rho^{\otimes N}$: (i)~the one presented here, where one fixes the extracted work, as  $w=(1-\gamma)N \mathcal{W}$, and observes an exponential decay of the failure probability, and (ii)~the single-shot regime, where one fixes a (small) failure probability, and  maximises the extracted work finding $w=N\mathcal{W}_{\rm th}(\rho)+\mathcal{O}(1/\sqrt{N})$ \cite{brandao:2013,horodecki2013fundamental,aaberg2013trulyNC,gour2015resource,chubb2017beyond}. Other choices of $\gamma$ lead to interesting interplays, such as \eqref{convergence},  between the success probability and the extracted work.  These different asymptotic scenarios are known as small, large, and moderate deviation regime in the context of information theory (see \cite{chubb2017moderate} and references therein).

Finally, let us briefly comment on the notion of work used here, based on a transition $0\rightarrow w$ in the state of W~\cite{brandao:2013,horodecki2013fundamental}. When  transitions between more energy levels of W are considered,  apparent violations of the second law can appear~\cite{Gemmer2015,gallego2016thermodynamic,ng2017surpassing}, which can be avoided by either accounting for the entropy increase in the state of W~\cite{Gemmer2015,gallego2016thermodynamic} or by restricting to protocols that act  on W in a translationally invariant manner~\cite{Popescu2013}. In this sense, we note that (i) since $\tilde{P}(w)\rightarrow 1$ the entropy gain can be neglected in most cases, and that (ii) it is possible to adapt the protocols derived here to satisfy translational invariance in W without modifying $\tilde{P}(w)$ \cite{Note6}. 
Another relevant question is the presence of quantum coherence in $\rho$. While the coherent part of $\mathcal{W}_{\rm th}(\rho)$ can not be extracted  for $U$'s that commute with the total Hamiltonian, it can  by considering global protocols on $\rho^{\otimes N}$~\cite{skrzypczyk2013extracting,lostaglio2015description,vaccaro2018coherence} --note that, since such protocols do not rely on B, they can  also be applied to $\mathcal{W}_{\rm erg}^{\rm glob}(\rho)$. This is yet another interesting effect of collective protocols.

\section{Conclusions} 
We have shown that collective operations can extremely reduce work fluctuations, which can decrease exponential with $N$ --the number of copies being processed. 
 This is in contrast to the standard ``thermodynamic limit", where the relative size of fluctuations decays slowly as $1/\sqrt{N}$. 
The exponential decay \eqref{boundII} of work fluctuations has been proven for generic systems both in the presence and absence of a thermal bath, providing a bound on the exponential decay for each case. This result contributes to our understanding  of global effects in quantum thermodynamics, 
and in this sense we note that previous results have shown that global operations can extract more average work~\cite{Oppenheim2002,Alicki2013,Frey2014,Frey2014,Raam2016,Mukherjee2016,Lekscha2018,
beny2017energy,gelbwaser2018cooperative,vitagliano2018trade} and in a faster manner~\cite{Karen2013,Binder2015,Campaioli2017,Ferraro2018,Thao2018,andolina2018charger,campaioli2018quantum}.
 
A high level of control has been assumed to obtain these results and arguably one of the most interesting questions is to which extent they can be observed experimentally. As recently proposed in Ref. \cite{Ferraro2018}, a promising candidate are architectures based upon the Dicke model \cite{dicke1954rh}, which allow for generating genuinely collective interactions. Another interesting proposal  is the one of Ref. \cite{lorch2018optimal} based on Josephson junctions, that would allow for testing various properties of work extraction protocols, including collective effects. These  proposals  open the door for experimental observations of some of the results reported here in near-future experiments. 

\section*{Acknowledgments} We thank Joseph M. Renes and Renato Renner   for useful discussions, and Kamil Korzekwa for valuable comments on the manuscript. M.P.-L. acknowledges support from the Alexander von Humboldt Foundation. This research was supported in part by the National Science Foundation under Grant No. NSF PHY-1748958.\\

\appendix

\section{Lower bounds on $\tilde{P}$ for $N$ qudits in thermally isolated systems}
\label{section.qdit}

We consider that S to be made up of $N$ identical qudit systems, with  a Hamiltonian $H_S= \sum_{i=1}^{N} \id^{\otimes i-1} \otimes h_S \otimes \id^{\otimes N-i+1}$ with $h_S= \sum_{i=1}^d \nu_i \ket{i}\bra{i}$, and an initial diagonal state  $\rho^{\otimes N}$ with $\rho=\sum_{i=1}^d p_i \ket{i}\bra{i}$.  The possible values of work are given by energy differences within S,
\begin{align}
w_{kl}= \sum_i (n_i-m_i) \nu_i,
\label{app.values.work}
\end{align}
where $n_i,m_i\in [0,N]$,  are natural numbers with $\sum_i n_i=\sum_i m_i=N$. We assume that the spectra $H_W$ of the work repository W system is dense enough so that it can accept the values of work \eqref{app.values.work}. One may take it to be continuous, $H_W = \int {\rm d}x \hspace{1mm}x \ket{x} \bra{x} $, but in principle this is not necessary. 
It  is convenient to introduce the vector notation ${\bf k}\equiv\{k_1,k_2,...,k_d \}$, 
\begin{align}
p({\bf k})=\prod_{i=1}^d p_i^{k_i}
\label{vecnot}
\end{align}
 and
\begin{align}
C^{{\bf k}}_N=\frac{N!}{\prod_{i=1}^d k_i!}
\label{vecnot2}
\end{align}
with $\sum_{i=1}^d k_i =N$. 

The average extractable work from $\rho$ (under unitary operations on S, but note also \cite{Note2}) is bounded by its ergotropy \cite{Allahverdyan2004},
\begin{align}
\mathcal{W}=  \sum_{i=1}^d (p_i - p_i^{\downarrow})\nu_i
\label{app.erg}
\end{align} 
where $p_i^{\downarrow}$ are the probabilities $p_i$ ordered in decreasing order, $p_{i+1}^{\downarrow}\leq p_i^{\downarrow}$,  $\forall i$. 
We wish to extract deterministically an amount of work $w=(1-\gamma)N\mathcal{W}$ by acting globally on $\rho^{\otimes N}$ and W. For that 
we take the $n_i$, $m_i$ in \eqref{app.values.work} to satisfy 
\begin{align}
k_i \equiv n_i-m_i
\label{n_iapp}
\end{align}
with 
\begin{align}
k_i = \floor{N (1-\gamma) (p_i-p_i^{\downarrow})},
\label{n_iII}
\end{align}
where the $\floor{}$ is introduced to ensure that the $n_i$ are natural numbers. In what follows we use that
\begin{align}
k_i \approx N (1-\gamma) (p_i-p_i^{\downarrow}) 
\label{n_i}
\end{align}
and assume \eqref{n_i} for the sake of mathematical tractability (when we consider numerical simulations  we use \eqref{n_iII}). 
It is important to recognize that there are many other choices of ${\bf n}$ that could lead to the desired $w=(1-\gamma)N\mathcal{W}$. For example, one can simply increase $n_1$ while taking $n_j=0$ with $j>1$. The choice \eqref{n_i} will however turn out to be crucial for the proof. 

We now consider the calculation of $\tilde{P}(w)$. Proceeding as in the main text, we have that at a global energy $E=\sum_i n_i \nu_i$ of SW:
\begin{itemize}
 \item Initial state ($\rho_W=\ket{0}\bra{0}$). There are $C^{{\bf n}}_N$ states with population $p({\bf k})$ for each ${\bf k}$.
 \item Target state ($\rho_W=\ket{w}\bra{w}$). There are $C_{N}^{{\bf n}-{\bf k}}$ states, all of them with zero population.
\end{itemize}
Then, since we can only move energies in degenerate levels, we obtain
\begin{align}
\tilde{P}(w)=\sum_{{\bf n}} \min\left(C^{{\bf n}}_N,C_{N}^{{\bf n}-{\bf k}} \right)p({\bf n}).
\label{appPsuc}
\end{align}
In what follows, we find $\min\left(C^{{\bf n}}_N,C_{N}^{{\bf n}-{\bf k}} \right)$ within the typical subspace of S (see, e.g. \cite{wilde2013quantum}).  To define the typical subspace, let $ \ket{i_1,...,i_N}$ with $i_j \in \{ 0,1,...,d\}$ be an energy eigenstate of S, and $n_l=\sum_{j=1}^{N}\delta_{i_j,l}$ the number of $l$'s in $ \ket{i_1,...,i_N}$.  The (strong)  $\delta$-typical subspace $\mathcal{T}^{\delta}_S$ is defined by eigenstates whose $n_i^{\rm typ}$  satisfy  
 $\big|p_i- n_i^{\rm typ}/N \big|\leq \delta$ $\forall i$ (and assuming $p_i \neq 0$). For convenience, let us also make the formal choice
\begin{align}
\delta=\gamma c
\label{deltafirstap}
\end{align}
where $c$ is still to be defined. Then, we can write a typical state as
\begin{align}
n_i^{\rm typ}=N(p_i+\gamma c_i)
\label{kityp}
\end{align}
with  $\sum_i c_i=0$ and $c_i \leq c$. Combining \eqref{kityp} with  \eqref{n_i} we obtain
\begin{align}
n_i^{\rm typ}-k_i=N(p_i^{\downarrow}+\gamma(c_i+p_i-p_i^{\downarrow})).
\end{align}
In order to find $\min\left(C^{{\bf n}^{\rm typ}}_N,C_{N}^{{\bf n}^{\rm typ}-{\bf k}} \right)$, we consider 
\begin{align}
\ln \frac{C_N^{{\bf n}^{\rm typ}-{\bf k}}}{C_N^{{\bf n}^{\rm typ}}} = & 
N\left(\sum_{i=1}^d \left(p_{i}+\gamma c_i \right) \ln[p_i+\gamma c_i]-\left(p_i^{\downarrow}+\gamma(c_i+p_i-p_i^{\downarrow})\right) \ln[p_i^{\downarrow}+\gamma(c_i+p_i-p_i^{\downarrow})]\right)
\nonumber\\
&+\frac{1}{2}\left(\sum_{i=1}^d  \ln[p_i+\gamma c_i]-\ln[p_i^{\downarrow}+\gamma(c_i+p_i-p_i^{\downarrow})]\right)+\mathcal{O}\left(\frac{1}{N}\right)
\label{appCNexp}
\end{align}
where we applied Stirling's approximation  $N! = \sqrt{2\pi N} N^N/e^N(1+\mathcal{O}(1/N))$. Expanding for small $\gamma$ we further obtain
\begin{align}
\ln \frac{C_N^{{\bf n}^{\rm typ}-{\bf k}}}{C_N^{{\bf n}^{\rm typ}}} = & N \gamma\left(S(\rho_{\rm pas}||\rho) -\sum_{i=1}^d c_i \ln \frac{p_i^{\downarrow}}{p_i} \right)+\gamma \frac{1}{2}\sum_i \left(\frac{c_i+p_i-p_i^{\downarrow}}{p_i^{\downarrow}}-\frac{c_i}{p_i} \right) +\mathcal{O}(\gamma^2)+\mathcal{O}\left(\frac{1}{N}\right)
\end{align}
where $\rho_{\rm pas}=\sum_i p^{\downarrow}_i \ket{i}\bra{i}$, $S(\rho_{\rm pas}||\rho)$ is the relative entropy, $S(\rho_{\rm pas}||\rho)=\sum_{i=1}^d (p_i^{\downarrow}-p_i)\ln p_i^{\downarrow}$. Let us write the $c_i$ as 
\begin{align}
c_i=c\alpha_i
\end{align}
with $\alpha_i \in [0,1]$,  $\sum_i \alpha_i=0$, and $c$ to be determined. Now, by noting that $S(\rho_{\rm pas}||\rho)\geq 0$ and  $\sum_i \left(\frac{p_i-p_i^{\downarrow}}{p_i^{\downarrow}}\right)\geq 0$, we can already anticipate that there is a sufficiently small positive $c$ such that $\ln (C_N^{{\bf n}^{\rm typ}-{\bf k}}/C_N^{{\bf n}^{\rm typ}}) \geq 0$ $\alpha_i$, and hence  $\min\left(C^{{\bf n}^{\rm typ}}_N,C_{N}^{{\bf n}^{\rm typ}-{\bf k}} \right)=C^{{\bf n}^{\rm typ}}_N$. We now want to find the maximum $c$ that ensures that $\ln (C_N^{{\bf n}^{\rm typ}-{\bf k}}/C_N^{{\bf n}^{\rm typ}}) \geq 0$ $\alpha_i$. This leads to
\begin{align}
c \leq \frac{S(\rho_{\rm pas}||\rho)+\frac{1}{2N}\left(\sum_{i=1}^d \frac{p_i-p_i^{\downarrow}}{p_i^{\downarrow}} \right)}{\sum_{i=1}^d \alpha_i \left(\ln \frac{p_i^{\downarrow}}{p_i} +\frac{1}{N}\left(\frac{1}{p_i}-\frac{1}{p_i^{\downarrow}} \right)\right)} +\mathcal{O}\left(\gamma\right)+\mathcal{O}\left(\frac{1}{N^2}\right)\hspace{3mm} \forall \alpha_i \hspace{2mm} {\rm s.t.} \hspace{2mm} \alpha_i \in [0,1], \hspace{1mm} \sum_i \alpha_i=0
\label{c_ibb}
\end{align}
In order to find a bound that is independent of the $\alpha_i$'s, we can maximise the denominator of \eqref{c_ibb}. For that, we can define the vector $h_i=p_i^{\downarrow}/p_i$, $i=1,...,d$ and order it decreasingly $h_i^{\downarrow}$ with $h_{i+1}^{\downarrow}\leq h_i^{\downarrow}$. Then, we have that
\begin{align}
\sum_{i=1}^d& \alpha_i \left(\ln \frac{p_i^{\downarrow}}{p_i} +\frac{1}{N}\left(\frac{1}{p_i}-\frac{1}{p_i^{\downarrow}} \right)\right) 
\nonumber\\
&=\sum_{i=1}^d \alpha_i \left(\ln \frac{p_i^{\downarrow}}{p_i} +\frac{1}{N}\left(\frac{\frac{p_i^{\downarrow}}{p_i}-1}{\frac{p_i^{\downarrow}}{p_i}} \right)\right) 
\nonumber\\
&\leq \sum_{i=1}^{\floor{d/2}} \left(\ln h_i^{\downarrow} +\frac{1}{N}\left(\frac{h_i^{\downarrow}-1}{h_i^{\downarrow}} \right)\right) - \sum_{i=\ceil{d/2}+1}^{d}\left(\ln h_i^{\downarrow} +\frac{1}{N^2}\left(\frac{h_i^{\downarrow}-1}{h_i^{\downarrow}} \right)\right) 
\end{align}
Defining, 
\begin{align}
\mathcal{C} \equiv \sum_{i=1}^{\floor{d/2}} \left(\ln h_i^{\downarrow} +\frac{1}{N}\left(\frac{h_i^{\downarrow}-1}{h_i^{\downarrow}} \right)\right) - \sum_{i=\ceil{d/2}+1}^{d}\left(\ln h_i^{\downarrow} +\frac{1}{N}\left(\frac{h_i^{\downarrow}-1}{h_i^{\downarrow}} \right)\right) ,
\end{align}
we finally arrive at the choice for $\delta$ \eqref{deltafirstap}
\begin{align}
\delta =  \frac{\gamma}{\mathcal{C} } \left(S(\rho_{\rm pas}||\rho)+\frac{1}{2N}\left(\sum_{i=1}^d \frac{p_i-p_i^{\downarrow}}{p_i^{\downarrow}} \right) +\mathcal{O}\left(\frac{1}{N^2}\right) \right) +\mathcal{O}\left(\gamma^2\right).
\label{appqudit}
\end{align}
Here we multiplied the term $\mathcal{O}(1/N^2)$ by $\gamma$ because for  $\gamma = 0$ we have that $C^{{\bf k}^{\rm typ}}_N = C_{N}^{{\bf k}^{\rm typ}-{\bf n}}$, and hence it must vanish.
Note that the choice \eqref{appqudit} only depends on $\gamma$ and on the initial state, defined by ${\bf p}^{\downarrow}$.  

Coming back to \eqref{appPsuc}, this result allows us to write,
\begin{align}
\tilde{P}(w)&=\sum_{{\bf n}} \min\left(C^{{\bf n}}_N,C_{N}^{{\bf n}-{\bf k}} \right)p({\bf n})
\nonumber\\
&\geq \sum_{{\bf n}\in \delta-{\rm typ}} \min\left(C^{{\bf n}}_N,C_{N}^{{\bf n}-{\bf k}} \right)p({\bf n})
\nonumber\\
&=\sum_{{\bf n}\in \delta-{\rm typ}} C^{{\bf n}}_N p({\bf n}).
\end{align}
Hence, all that is left is to evaluate the population in the $\delta$-typical subspace, i.e. the probability that $ |n_i/N-p_i|\leq \delta $,  $\forall i$. For that, we note that
 \begin{align}
\tilde{P}(w)&\geq  P\left(\left\{\left| \frac{n_i}{N}-p_i  \right| < \delta \right\}_i\right)
\nonumber\\ 
&=1-P\left(\bigcup_i \left| \frac{n_i}{N}-p_i  \right| \geq \delta \right) 
 \nonumber\\
 &\geq 1-\sum_{i=1}^d P\left( \left| \frac{n_i}{N}-p_i  \right|\geq\delta \right)
 \nonumber\\
 &\geq 1-d_S e^{-2N\delta^2}
 \label{derivationqudits}
\end{align}
where in the first inequality we used union bound of probability theory, and in the second  inequality we used the Hoeffding's inequality. Because $\delta$ is proportional to $\gamma$ as in \eqref{appqudit}, we obtain the desired decay in $N\gamma^2$ with corrections of order $N\mathcal{O}(\gamma^3)$ and $\frac{1}{N}\mathcal{O}(\gamma^2)$.  Absorbing the factor 2 into $c$, we obtained the desired result from the main text.

\section{Passivity, complete passivity, and work fluctuations.}
\label{section.fluctuations.activation}
In this section, we extend our previous results when \eqref{app.erg} is substituted by the global bound 
\begin{align}
\mathcal{W}_{\rm erg}^{\rm (glob.)}(\rho) \equiv \lim_{N \rightarrow \infty} \frac{\mathcal{W}_{\rm erg}(\rho^{\otimes N})}{N}= {\rm Tr}(h(\rho-\omega_{\beta}(h))),
\label{thermoBound.app}
\end{align}
where $\omega_{\beta}(h)$ is a thermal state, $\omega_{\beta}(h)\equiv e^{-\beta h}/{\rm Tr}(e^{-\beta h})$, whose temperature is defined by $S(\rho)=S(\omega_{\beta}(h))$ with $S(\rho)=-{\rm Tr}(\rho \ln \rho)$ the Von Neumann entropy. It is convenient to expand  $\omega_{\beta}(h)$,
\begin{align}
\omega_{\beta}(h)=\sum_i p_i^{\rm th} \ket{i} \bra{i}
\end{align}
In order to extend our previous considerations, the key idea is to chose the $n_i$ in \eqref{n_iapp} as
\begin{align}
k_i = \floor{N (1-\gamma) (p_i-p_i^{\rm th.})},
\label{an_iII}
\end{align}
and again we assume
\begin{align}
k_i \approx N (1-\gamma) (p_i-p_i^{\rm th.}), 
\label{an_i}
\end{align}
for the sake of analytical tractability. Proceeding as in the previous section, one then obtains,
\begin{align}
\frac{1}{N}\ln \frac{C_N^{{\bf n}^{\rm typ}-{\bf k}}}{C_N^{{\bf n}^{\rm typ}}} =  
\sum_{i=1}^d \left(p_{i}+\gamma c_i \right) \ln[p_i+\gamma c_i]-\left(p_i^{\rm th.}+\gamma(c_i+p_i-p_i^{\rm th.}) \ln[p_i^{\rm th.}+\gamma(c_i+p_i-p_i^{\rm th.})]\right)+\mathcal{O}\left(\frac{1}{N}\right)
\label{appCNexp}
\end{align}
where we now keep only highest orders in $N$. Expanding over $\gamma$ as before, 
\begin{align}
\frac{1}{N}\ln \frac{C_N^{{\bf n}^{\rm typ}-{\bf k}}}{C_N^{{\bf n}^{\rm typ}}} = & \gamma\left(S(\omega_{\beta}(h)||\rho) -\sum_{i=1}^d c_i \ln \frac{p_i^{\rm th.}}{p_i} \right) +\mathcal{O}(\gamma^2)+\mathcal{O}\left(\frac{1}{N}\right)
\end{align}
where we used that $S(\rho)=S(\omega_{\beta}(h))$. At this point, the proof proceeds in complete analogy with the previous one, so one easily arrives at the final result
\begin{align}
c=\frac{\sqrt{2}\hspace{0.5mm}S(\omega_{\beta}(h)||\rho)}{\sum_{i=1}^{\floor{d/2}} \ln h_i^{\downarrow} - \sum_{i=\ceil{d/2}+1}^{d} \ln h_i^{\downarrow} }
\end{align}
with $h_i=p_i^{\rm th}/p_i$, $i=1,...,d$.


\section{Lower bounds on $\tilde{P}$ in the presence of a bath}
\label{section.qdit.bath} 
Let us start by making explaining   $H_B$  following  \cite{horodecki2013fundamental}.  Denote by $e_B$ an energy of B and by $g(e_B)$ its degeneracy, and recall that B is assumed to be in a Gibbs state $\omega_{\beta}(H_B)$. Then,   we assume that there exists a subset $\mathcal{E}$ of the whole spectrum $\{ e_B \}$, in which $g_B(e_B)$ increases exponentially and satisfies $g_B(e_B-e_S)=g_B(e_b)e^{-\beta e_S}$, where $e_S$ is any energy  of $S$-- see \cite{horodecki2013fundamental} for more details. For short-range interacting  systems, the subset $\mathcal{E}$ contains $1-\epsilon$ of the population of $\omega_{\beta}(H_B)$, with $\epsilon$ approaching 0 with the size of B \cite{gemmer2009quantum}. Therefore, in what follows we assume that B is sufficiently large so that $\epsilon \approx 0$, and hence $H_B$ is well described by the subset $\mathcal{E}$. 
 

Having defined B, we now proceed as in the last section and consider work extraction from  $\rho^{\otimes N}$,  with   $\rho=\sum_{i=1}^d p_i \ket{i}\bra{i}$ and $H_S= \sum_{i=1}^d \nu_i \ket{i}\bra{i}$. The initial state of SBW being $\rho^{\otimes N} \otimes \omega_{\beta}(H_B) \otimes \ket{0} \bra{0}$.   Finding  $\tilde{P}(w)$ essentially boils down to moving states in the degenerate energy levels of SBW. Let us fix a total energy $E$ for SBW and note that (using the notation \eqref{vecnot} and \eqref{vecnot2}), 
\begin{itemize}
\item  Initial state. Given $W$ at $\ket{0}$, $e_S=\sum_i n_i \nu_i$ and $e_B=E-\sum_i n_i \nu_i $, there are $g(E)e^{-\beta \sum_i n_i \nu_i}C_{N}^{\bf k}$ states at total energy $E$, each with probability $\mathcal{Z}^{-1}e^{-\beta E} e^{\beta \sum_i n_i \nu_i} p({\bf n})$ and $\forall {\bf n}$.
\item Target state. Given $W$ at $\ket{w}$, $e_S=\sum_i n_i \nu_i$ and $e_B=E-\sum_i n_i \nu_i -w $, there are $g(E)e^{-\beta \omega}e^{-\beta \sum_i n_i \nu_i}C_{N}^{\bf n}$ states at total energy $E$ (each of them initially with 0 probability), and $\forall {\bf n}$.
\end{itemize}
We now wish to move the  probabilities of the initial to the target state, which has $e^{-\beta w}$ less states, in each degenerate energy level $E$. 
Since the scenario is identical for each global energy, we can focus on a particular $E$ with a degeneracy $g(E)=g$, which we take to be a very large number.
 By normalising the probabilities within this subspace, we can reformulate the setting as: 
 In a subspace of SBW with fixed global energy, the initial state of SBW  has normalised populations of the form 
\begin{align}
p_{\rm ini.}({\bf n})=g^{-1}e^{\beta \sum_i n_i \nu_i} p({\bf n})
\end{align} 
with degeneracies 
\begin{align}
g_{\rm ini.}({\bf n})=ge^{-\beta \sum_i n_i \nu_i}C^{{\bf n}}_N,
\end{align}
 with $C^{{\bf n}}_N=N!/\prod_{i} n_i!$ and   $n_i=0,..,N$ with $\sum_i n_i=N$. On the other hand, in the target state, where $W$ is at $\omega$, has degeneracies 
 \begin{align}
 g_{\rm tar}({\bf n})=ge^{-\beta (\omega+ \sum_i n_i \nu_i)}C^{{\bf n}}_N.
 \end{align}
  The total number of states in the target state is then given by
\begin{align}
\mathcal{N}_{\rm tar}&=ge^{-\beta \omega} \sum_{{\bf n}}  e^{-\beta \sum_i n_i \nu_i} C^{{\bf n}}_N=ge^{-\beta \omega}\left(\sum_i e^{-\beta \nu_i}\right)^{N}= g e^{-\beta \big(w+ N F(\omega_{\beta}(H_S))\big)}.
\label{total.des.II}
\end{align}
To find $\tilde{P}(w)$, in principle one has to move the $\mathcal{N}_{\rm tar}$ biggest populations of the initial state to the target state. This leads to a set of inequalities that are directly related to the concept of thermomajorisation \cite{horodecki2013fundamental}. In order to obtain a simple and  analytic result,  here we will instead move the typical subspace of S, hence finding a lower bound on $\tilde{P}(w)$. 
Recall that the (strong)  $\delta$-typical subspace $\mathcal{T}^{\delta}_S$ is defined by eigenstates that satisfy  $n_l \in  N[p-\delta,p+\delta]$, $\forall l$ with $\delta>0$ where  $n_l=\sum_{j=1}^{N}\delta_{i_j,l}$ is the number of $l$'s in the state $ \ket{i_1,...,i_N}$.
The number of states in $\mathcal{T}^{\delta}_S$ can be upper bounded by $e^{N(S(\rho)+C\delta)}$ with $C=-\sum_i \ln p_i$ for $p_i\in (0,1)$ \cite{wilde2013quantum}. This allows us to upper bound the number of typical states of SBW at global energy $E$ in the initial state,
\begin{align}
\mathcal{N}_{\rm typ.ini.}^{(\delta)} &=  \sum_{{\bf n}\in {\rm typ}} g_{\rm ini.}({\bf n})  
\leq g e^{-\beta N \sum_i \nu_i(p_i-\delta)}   \sum_{{\bf n}\in {\rm typ}}  C^{{\bf n}}_N
  \leq g e^{-\beta N \sum_i \nu_i(p_i-\delta)+N(S(\rho)+C\delta) }
  \nonumber\\
  &=ge^{-\beta N F(\rho)+\delta N (\beta \sum_i \nu_i-\sum_i \ln p_i)}.
\label{total.ini.II}
\end{align}
The goal now is to find the largest $\delta$ that guarantees $\mathcal{N}_{\rm tar} \geq \mathcal{N}_{\rm typ.ini.}^{(\delta)}$, so that we can move the whole population of the  typical space to the target space. This easily leads to
\begin{align}
\delta=\frac{\beta \gamma \Delta F}{\beta \sum_i \nu_i -\sum_i \ln p_i}.
\label{delta.app}
\end{align}
 Putting everyting together, and proceeding exactly as in \eqref{derivationqudits},  we obtain
\begin{align}
\tilde{P}(w) \geq 1-d\exp\left(-N\gamma^2 \left( \frac{\sqrt{2}\beta  \mathcal{W}_{\rm th}(\rho)}{\beta \sum_i \nu_i -\sum_i \ln p_i}\right)^2\right)
\label{Pwfib}
\end{align}
with $w=N \mathcal{W}_{\rm th}(\rho)(1-\gamma)$, and where we used that $\mathcal{W}_{\rm th}(\rho)=F(\rho)-F(\omega_{\beta}(h))$. This shows the desired exponential decay with $N\gamma^2$, which is multiplied by a function that depends on $\beta$ and on $S$ through $H_S$ and $\rho_S$.

\section{Numerical computation of $\tilde{P}(w)$}
\label{appnumerics}

Here we  discuss how to compute $\tilde{P}(w)$ exactly by numerical means.  First of all, in the case of thermally isolated systems, because of the high degeneracies, we can describe the state effectively with a vector of dimension $\approx N$. Then, we can compute $\tilde{P}(w)$ by computing exactly \eqref{eq: P_exact_binom}. 

\begin{figure}[htp]
\centering
\includegraphics[scale=0.3]{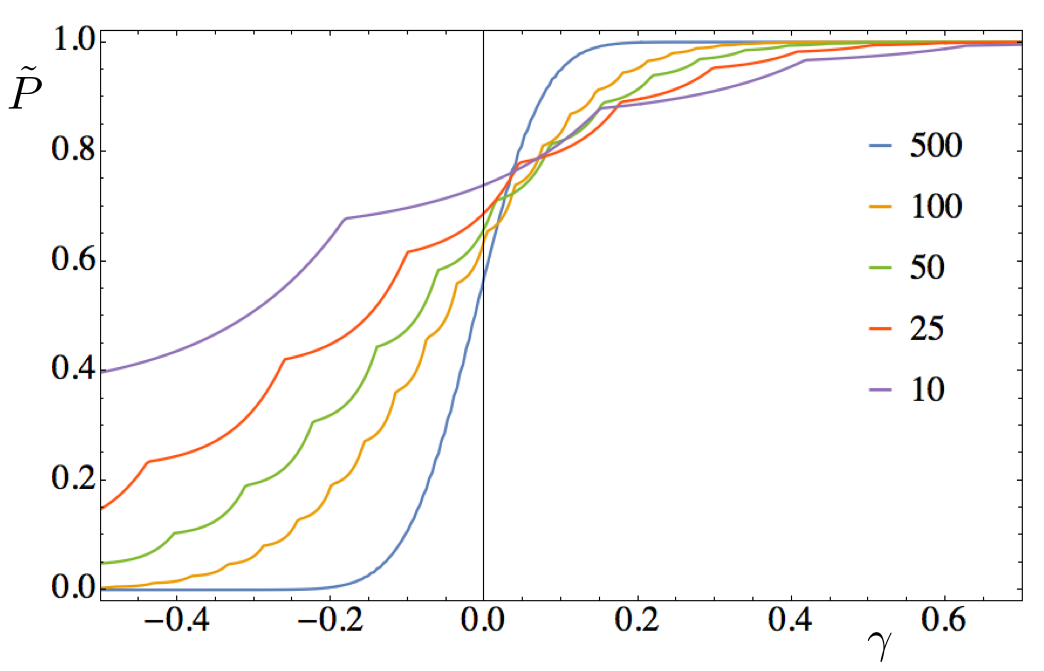}
\caption{\label{fig:resultxx}   Deterministic work extraction in the presence of a bath. $\tilde{P}(\omega_{\gamma})$ as a function of $\gamma$, with $\omega_{\gamma}=(1+\gamma)N\mathcal{W}_{\rm th}(\rho)$ and where $\mathcal{W}_{\rm th}(\rho)$ is the free energy difference.  Parameters: $p=0.8$, $\beta=1$, $\nu=1$, $g=2^{10^{4}}$. }
\end{figure}

The exact numerical calculation of $\tilde{P}(w)$ when the thermal bath is present is more involved. 
 Essentially, to compute $\tilde{P}(w)$ we want to move as much probability as possible from the initial state to the target state at each energy level $E$ (See Section \ref{section.qdit.bath}). As explained in Sec. \ref{section.qdit.bath}, we recall that it is enough to focus on a single energy level E.  In particular, we need to move the   $\mathcal{N}_{\rm tar}$ (given in \eqref{total.des.II}) biggest populations of the initial state to the target state. We do so numerically for the case of qubits with population $p$. The populations of the initial state can be written as 
 \begin{align}
 p_{\rm ini}(k)=g^{-1}e^{\beta k \nu} p^k (1-p)^{N-k}=g^{-1}e^{k(\beta \nu+\ln p -\ln(1-p))+n\ln(1-p)}
 \label{pini}
 \end{align}
 with degeneracy
\begin{align}
g_{\rm ini}(k)=ge^{-\beta k \nu}C^{k}_N.
\end{align} 
From \eqref{pini}, we see that the probability decreases/increases monotonically with $k$ if $\beta \nu+\ln p -\ln(1-p)$ is negative/positive. Hence, we can construct a simple algorithm as follows:
\begin{enumerate}
\item Choose a large number for $g$ (in Fig. \ref{fig:resultxx} we take $g=2^{10^4}$), which approximately sets the dimension of B. If $g$ is large enough, the results become independent of it. 
\item Check if $\beta \nu+\ln p -\ln(1-p)$ is negative/positive.
\item If it is negative: 
\begin{itemize}
\item for $k=0$, fill $g_{\rm ini}(0)$ states of the target state with probability $p_{in}(0)$. Increase $k$ by 1.
 \item for $k=1$,  fill $g_{\rm ini}(1)$ states of the target state with probability $p_{in}(1)$.  Increase $k$ by 1.
 \item Idem for larger $k$'s until all states of the target state are filled, i.e., one reaches $\mathcal{N}_{\rm tar}$ given in \eqref{total.des.II}.
\end{itemize}  
\item If it is positive: 
\begin{itemize}
\item for $k=k$, fill $g_{\rm ini}(k)$ states of the target state with probability $p_{\rm ini}(N)$. Decrease $k$ by 1.
 \item for $k=N-1$,  fill $d_{\rm ini}(N-1)$ states of the target state with probability $p_{\rm ini}(N-1)$. Decrease $k$ by 1.
 \item Idem for smaller $k$'s until all states of the target state are filled.
\end{itemize}  
\end{enumerate}
This protocol ensures that as much probability as possible has been transferred to the target state. Results for $p=0.8$ are shown in Fig. \eqref{fig:resultxx}.

\end{document}